\documentclass[pdflatex,sn-nature]{sn-jnl}

\usepackage{graphicx}
\usepackage{multirow}
\usepackage{amsmath,amssymb,amsfonts}
\usepackage{xcolor}
\usepackage{booktabs}
\usepackage{siunitx}
\unnumbered  

\newcommand{\Win}{W_{\mathrm{in}}}
\newcommand{\WinR}{W_{\mathrm{in}}^{(r)}}
\newcommand{\dTfW}{\delta T_{\mathrm{fW}}}
\newcommand{\Trta}{T_{\mathrm{RTA}}}
\newcommand{\TfW}{T_{\mathrm{fW}}}
\newcommand{\keff}{k_{\mathrm{eff}}}
\newcommand{\kRTA}{k_{\mathrm{RTA}}}
\newcommand{\kdense}{k_{\mathrm{dense}}}
\newcommand{\kbulk}{k_{\mathrm{bulk}}}
\newcommand{\Ma}{M_{a}}
\newcommand{\yes}{$\checkmark$}
\newcommand{\no}{$\times$}

\begin{document}


\title{Full-Scattering-Matrix Deterministic Phonon Boltzmann Transport Simulation}

\author*[1]{\fnm{Y. Sungtaek} \sur{Ju}}\email{sungtaek.ju@ucla.edu}

\affil*[1]{\orgdiv{Department of Mechanical and Aerospace Engineering}, \orgname{UCLA},
  \orgaddress{\street{420 Westwood Plaza}, \city{Los Angeles}, \state{CA},
    \postcode{90025}, \country{U.S.A.}}}

\abstract{%
Solutions to the phonon Boltzmann transport equation under the relaxation-time approximation (RTA) are fundamentally limited in that they do not account for the off-diagonal elements of the scattering matrix, which encode intermode energy redistribution. We find that the phonon in-scattering matrix is globally incompressible, requiring nearly its full rank for any useful Frobenius accuracy.  The incompressibility worsens as the Brillouin zone is refined.  We show that, despite this difficulty, one can develop a computationally efficient 3D BTE solver incorporating the complete scattering matrix by leveraging our two structural discoveries:  the non-equilibrium phonon distribution inhabits a remarkably low-dimensional subspace of mode space regardless of how many phonon modes exist, and the leading singular modes of the scattering operator align selectively with this transport-active subspace. Consequently, truncation incurs negligible transport error even under large norm error. The phonon streaming operator's mode-diagonal character further motivates a hybrid architecture that exploits these two properties. When applied to nanoscale structures emulating a fin field-effect transistor, our BTE solver quantifies a geometry-independent multiplicative correction to the temperature rise predicted under RTA. Our theoretical work and resulting BTE solver help enable rigorous study of phonon transport and systematic design of devices and structures in the ballistic and quasi-ballistic phonon transport regime.}


\maketitle


\section{Introduction}\label{sec:intro}

As semiconductor structures and devices continue to scale down, non-diffusive heat
transport is becoming increasingly significant. Experimental evidence that Fourier's law breaks down in sub-micron silicon even at room temperature was first provided by Ju and Goodson \cite{Ju1999}, who measured the thermal conductivity of silicon-on-insulator (SOI) thin films and found suppression consistent with phonon mean free paths being comparable to film thickness. This size effect was further confirmed using silicon nanowires by Li et al. \cite{Li2003}. These and many follow-up studies established that ballistic transport governs heat conduction at device-relevant scales in many important semiconductors.

The phonon BTE \cite{Majumdar1993} in steady state is
\begin{equation}
  \mathbf{v}_\lambda \cdot \nabla f_\lambda(\mathbf{r})
    = -\sum_{\lambda'} W_{\lambda\lambda'} f_{\lambda'}(\mathbf{r})
    + \dot{Q}_\lambda(\mathbf{r}),
  \label{eq:bte}
\end{equation}
where $f_\lambda$ is the phonon distribution function for mode
$\lambda = (\mathbf{q}, s)$, $\mathbf{v}_\lambda$ is the group velocity,
$\dot{Q}_\lambda$ is a volumetric heat source, and $W$ is the scattering matrix
encoding all three-phonon Normal and Umklapp processes, computable from interatomic
force constants \cite{ShengBTE,Phoebe}.

Three communities have made complementary progress on Eq.~(\ref{eq:bte}), but none
has achieved the combination of full $W$, three-dimensional geometry, and
deterministic finite-volume discretisation required for quantitative analysis of devices and structures. Table~\ref{tab:priorwork} summarises the prior work context.

The \textit{ab initio transport community} \cite{ShengBTE,Phoebe,Cepellotti2016,Simoncelli2020} generates the full $W$ and has established that off-diagonal scattering measurably affects phonon
transport \cite{Chiloyan2021}. These methods, however, are restricted to periodic bulk geometries and cannot treat finite-domain, source-driven boundary value problems.

The \textit{device BTE community} \cite{Sverdrup2001,Narumanchi2004,Hu2024,Shang2025} (see also physics-informed neural network approaches \cite{LiWang2022,ZhouLuo2023}) has developed methods for solving the BTE in 2D/3D geometries with realistic phonon dispersions and boundary conditions.
Existing solvers, however, employ the relaxation time approximation (RTA), replacing $W_{\lambda\lambda'}$ with $-\delta_{\lambda\lambda'}/\tau_\lambda$ and ignoring  intermode scattering
correlations. A recent review explicitly identifies reconciling RTA-based ballistic and hydrodynamic
formulations as an outstanding necessity for thermal management in nanodevices \cite{Beardo2025}.

The \textit{beyond-RTA Monte Carlo community} \cite{Landon2014,LiLee2019,Souza2023} demonstrated that full-$W$ BTE in finite domains differs from RTA --- with errors reaching 10--30\% at moderate Knudsen numbers in 2D materials and graphene ribbons. Monte Carlo methods, however, typically do not scale tractably to 3D structures or devices with the full phonon spectrum.

At first glance, the computational challenge appears straightforward: the $O(N_\lambda^2)$ cost of
$W$ suggests truncation based on SVD (singular value decomposition)  as the enabling technology, paralleling successful low-rank approximations in quantum chemistry and radiative transfer. We show, however, that this approach fails fundamentally for phonon-phonon scattering. The in-scattering operator $\Win$ requires 87--91\% of its full SVD rank for even 1\% Frobenius accuracy, and the required fraction worsens as the Brillouin zone is refined.

Conventional tensor-train compression along the full 6D mode-space indices likewise fails: we implemented and validated a complete TT-AMEn (tensor train alternating minimal energy) solver, which proved orders of magnitude slower than dense sweeps because the streaming operator is diagonal in mode space, making tensor-train coupling an overhead without computational benefit.

We also resolve why, despite the incompressibility of the scattering matrix, SVD truncation can still yield accurate solutions to the BTE, and how a computationally efficient solver can be built. We show that the non-equilibrium BTE solution occupies a low-rank subspace of mode space.  In 1D cases, two basis vectors describe the entire anisotropic distribution regardless of phonon mode count or device length. The leading SVD modes of $\Win$ are transport-selective: they align with this rank-2 subspace with selectivity 60--385$\times$, so that the almost 90\% of the SVD spectrum that is discarded acts entirely within the equilibrium mode subspace and has no effect on transport observables. This is reminiscent of the two-fluid phonon transport model \cite{JuGoodson1999} where phonons are grouped into the ``propagating'' mode or the ``reservoir'' mode.  Similar observations are made in 3D cases.

The end result is the first deterministic 3D BTE solver incorporating the full scattering matrix. When applied to a FinFET (fin field-effect transistor)-like structure, it quantifies the full-$W$ correction to be $11.2 \pm 0.3\%$ of the RTA temperature rise --- converged, geometry-independent, and physically interpretable.

\begin{sidewaystable}
\caption{Scope of prior work. All binary assessments reflect the scope of the
  cited work as reported. The final column records whether the method solves the
  phonon BTE with the complete off-diagonal scattering matrix for a finite
  geometry --- the combination this work achieves for the first time in three
  dimensions.}\label{tab:priorwork}
\small
\begin{tabular}{@{}p{5.2cm}lllll@{}}
\toprule
Method & Full $W$? & Phys.\ BCs? & 3D Si? & Det.\ FVM?
  & Full $W$, finite geom.? \\
\midrule
ShengBTE/Phoebe/relaxons\cite{ShengBTE,Phoebe,Cepellotti2016}
  & \yes & \no\ (periodic) & \no & \yes & \no\ (bulk only) \\
Chiloyan et al.\ (2021)\cite{Chiloyan2021}
  & \yes & \no\ (unbounded) & \no & \yes & \no\ (unbounded) \\
BTE-Barna (2023)\cite{Souza2023}
  & \yes & \yes\ (2D) & \no & \no\ (MC) & \no\ (2D only) \\
Landon \& Hadjiconstantinou (2014)\cite{Landon2014}
  & \yes & \yes\ (2D) & \no & \no\ (MC) & \no\ (2D only) \\
Li \& Lee (2019)\cite{LiLee2019}
  & \yes & \yes\ (2D) & \no & \no\ (MC) & \no\ (2D only) \\
Beardo et al.\ (2025)\cite{Beardo2025}
  & \no\ (RTA) & \yes & \yes & \yes\ (FEM) & \no\ (RTA only) \\
GiftBTE (2024)\cite{Hu2024}
  & \no\ (RTA) & \yes & \yes & \yes & \no\ (RTA only) \\
JAX-BTE (2025)\cite{Shang2025}
  & \no\ (RTA) & \yes & \yes & \yes & \no\ (RTA only) \\
\midrule
\textbf{This work}
  & \textbf{\yes} & \textbf{\yes} & \textbf{\yes} & \textbf{\yes}
  & \textbf{\yes} \\
\botrule
\end{tabular}
\footnotetext{Abbreviations: Phys.\ BCs~=~physical boundary conditions in a
  finite domain; Det.\ FVM~=~deterministic finite-volume method;
  MC~=~Monte Carlo; FEM~=~finite-element method.}
\end{sidewaystable}


\section{Results}\label{sec:results}

\subsection{Part I: Structure of the scattering operator}\label{sec:partI}

\subsubsection{Scattering channel count and matrix density}\label{sec:channels}

The off-diagonal entry $[\Win]_{\lambda\lambda'}$ is nonzero when a third mode
$\lambda''$ simultaneously satisfies crystal-momentum conservation
($\mathbf{q} \pm \mathbf{q}' = \mathbf{q}'' + \mathbf{G}$) and energy conservation
($\omega_\lambda \pm \omega_{\lambda'} = \omega_{\lambda''}$) on the Brillouin zone
grid. For an $N^3$ Monkhorst--Pack grid with $\Ma \approx 6N^3$ thermally active modes, the
number of valid scattering triplets per mode grows as $O(N^3)=O(\Ma)$, giving a total
channel count proportional to $\Ma^2$ --- the defining signature of a dense matrix.
Table~\ref{tab:channels} confirms this empirically: at $N = 9$ ($\Ma = 4371$), the
scattering matrix builder enumerates 35.8 million processes. Fitting the channel count to $\Ma^\alpha$ gives $\alpha = 1.946 \approx 2$.

\begin{table}[ht]
\caption{Scattering channel count confirms $O(\Ma^2)$ density.}\label{tab:channels}
\begin{tabular}{@{}rrrrr@{}}
\toprule
$N$ & $\Ma$ & Coalescence + Decay & $\Ma^2$ & Ratio \\
\midrule
 3 &    159 &        56,638 &    25,281 & 2.24 \\
 5 &    747 &     1,026,726 &   558,009 & 1.84 \\
 7 &  2,055 &     7,833,593 & 4,223,025 & 1.86 \\
 9 &  4,371 &    35,840,591 & 19,105,641 & 1.88 \\
\botrule
\end{tabular}
\end{table}

The ratio exceeds 1 because each scattering event involves three modes, giving roughly 2 matrix entries per triplet \cite{Ziman1960}. At all $N$, over 99\% of entries of $\Win$ exceed a relative threshold of $10^{-4}$ (Table~\ref{tab:svdincomp}), confirming that $\Win$ is fully dense at all BZ grid sizes studied.

\subsubsection{SVD rank requirement: global incompressibility}\label{sec:incomp}

\begin{figure}[ht]
\centering
\includegraphics{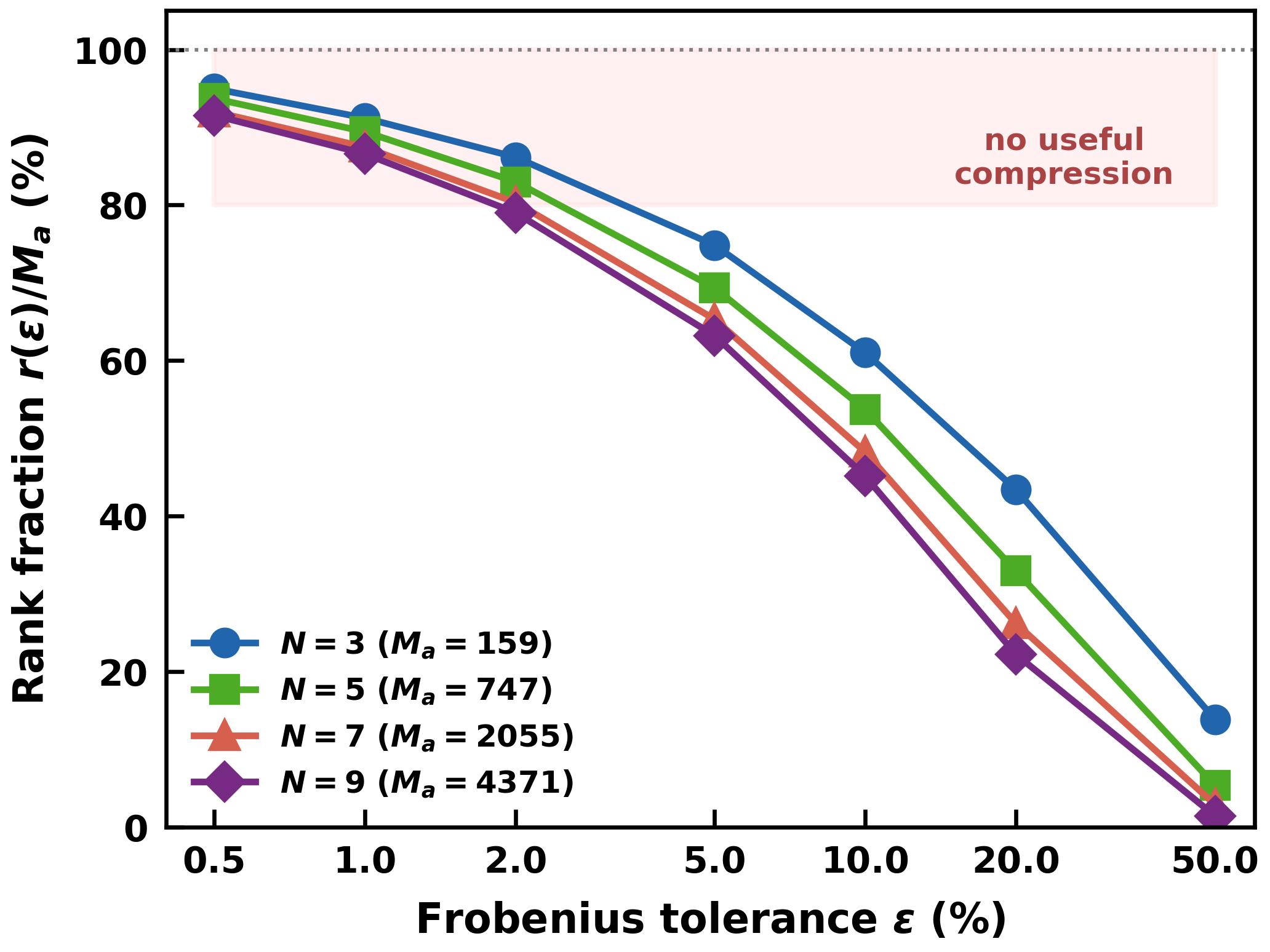}
\caption{SVD incompressibility of $\Win$. Rank fraction $r(\varepsilon)/\Ma$
  required to achieve Frobenius tolerance $\varepsilon$, for $N = 3, 5, 7, 9$.
  The shaded region ($r/\Ma > 80\%$) marks the regime with no useful compression.
  At $\varepsilon = 1\%$, the required fraction is 87--91\% at all grid sizes.}\label{fig:incompressibility}
\end{figure}

We compute the full SVD of $\Win$ at $N = 3, 5, 7, 9$ and measure the rank fraction $r(\varepsilon)/\Ma$ --- the minimum fraction of the full SVD rank needed to achieve Frobenius tolerance $\varepsilon$:
\begin{equation}
  \frac{\|\Win - \WinR\|_F}{\|\Win\|_F} \leq \varepsilon.
  \label{eq:frob}
\end{equation}
The results are shown in Table~\ref{tab:svdincomp} and Fig.~\ref{fig:incompressibility}. At $\varepsilon = 1\%$, the rank fraction is 91.2\% ($N=3$), 89.4\% ($N=5$), 87.4\% ($N=7$), and 86.6\% ($N=9$), respectively. The trend is a slow decrease: fitting $r(1\%)/\Ma$ against $\log_{10}(\Ma)$ gives a
slope of $-3.3\%$ per decade, extrapolating to 83\% at the production ab initio grid size $N = 20$ ($\Ma \sim 50{,}000$). The fraction never approaches zero.

\begin{table}[ht]
\caption{SVD incompressibility and density of $\Win$.}\label{tab:svdincomp}
\begin{tabular}{@{}rrrrrrrr@{}}
\toprule
$N$ & $\Ma$ & $r(0.5\%)$ & $r(1\%)$ & $r(5\%)$ & $r(10\%)$
    & $r(1\%)/\Ma$ & nnz${}_{10^{-4}}/\Ma^2$ \\
\midrule
 3 &    159 &   151 &   145 &   119 &   97 & 91.2\% & 99.1\% \\
 5 &    747 &   700 &   668 &   518 &  401 & 89.4\% & 99.5\% \\
 7 &  2,055 & 1,891 & 1,797 & 1,342 &  991 & 87.4\% & 99.2\% \\
 9 &  4,371 & 4,001 & 3,785 & 2,762 & 1,972 & 86.6\% & 99.1\% \\
\botrule
\end{tabular}
\end{table}

\begin{figure}[ht]
\centering
\includegraphics{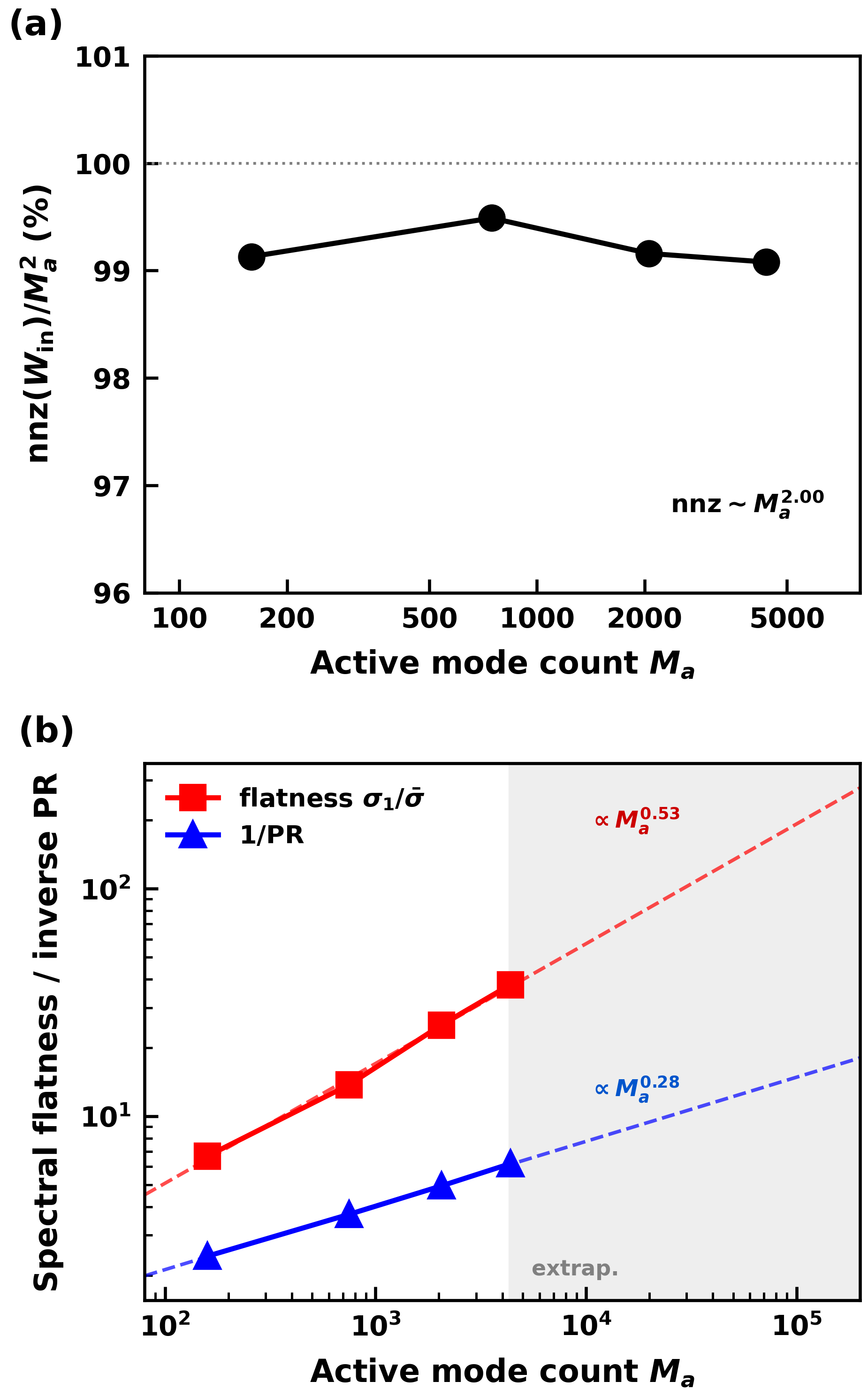}
\caption{Spectral density and flatness scaling.
  (a) Non-zero entry fraction $\mathrm{nnz}(\Win)/\Ma^2$ at threshold $10^{-4}$,
  showing $> 99\%$ density with power-law $\mathrm{nnz} \propto \Ma^{2.00}$
  ($R^2 = 1.000$).
  (b) Spectral flatness $\sigma_1/\bar\sigma$ and $1/\mathrm{PR}$ vs.\ $\Ma$
  on log--log axes, growing as $\Ma^{0.53}$ and $\Ma^{0.28}$ respectively.}\label{fig:flatness}
\end{figure}

The spectral flatness $\sigma_1/\bar{\sigma}$ (ratio of largest to RMS singular value)
grows as $\Ma^{0.53}$ --- the spectrum becomes progressively more uniform as the BZ is refined (Fig.~\ref{fig:flatness}). The participation ratio $\mathrm{PR} = (\sum_i \sigma_i)^2/(\Ma \sum_i \sigma_i^2)$ decreases as $\Ma^{-0.28}$, confirming no small subset of singular vectors captures a
disproportionate share of the Frobenius norm. The Eckart--Young theorem \cite{EckartYoung1936} guarantees these rank estimates are optimal. This shows that $\Win$ is globally incompressible via any low-rank representation at all physically relevant BZ grid sizes, and this incompressibility worsens with BZ refinement.

\subsubsection{The relaxon spectrum is gapless}\label{sec:relaxons}

An alternative compression route suggested by the relaxon theory \cite{Cepellotti2016,Simoncelli2020} is to diagonalise the symmetrised
scattering matrix $\tilde{W} = C^{1/2}WC^{-1/2}$, where $C = \mathrm{diag}(c_\lambda)$ is the heat-capacity matrix. If the eigenvalue spectrum had a gap between a small number of slow modes
($|\lambda_s| \ll |\lambda_{\max}|$) and the fast bulk, those slow modes would define
a compressible transport subspace. Table~\ref{tab:relaxons} shows the gap ratio $|\lambda_1|/|\lambda_{\Ma}|$: values are approximately 0.002--0.004 at all $N$, indicating an essentially gapless spectrum. The number of relaxons within 5\% of $|\lambda_{\max}|$ grows as $\Ma^{1.1}$ ---
approximately proportional to $\Ma$ --- providing no compression.

\begin{table}[ht]
\caption{Relaxon eigenvalue spectrum: gapless at all grid sizes.}\label{tab:relaxons}
\begin{tabular}{@{}rrrrrr@{}}
\toprule
$N$ & $\Ma$ & Gap ratio $|\lambda_1|/|\lambda_{\Ma}|$
    & Slow modes (1\%) & Slow modes (5\%) & Slow modes (10\%) \\
\midrule
 3 &    159 & 0.00324 &  1 &   3 &    9 \\
 5 &    747 & 0.00315 &  2 &   2 &   18 \\
 7 &  2,055 & 0.00236 &  2 &  43 &  810 \\
 9 &  4,371 & 0.00409 &  2 &  55 & 1,552 \\
\botrule
\end{tabular}
\end{table}

\subsubsection{Why tensor-train full-6D compression also fails: streaming
  diagonality}\label{sec:tt}

A full six-dimensional TT-AMEn solver on the index set $(\lambda, \theta, \varphi, x, y, z)$ was implemented and validated: residual $6.6 \times 10^{-16}$ vs.\ the dense reference, 99.2\% parametric compression, TT bond ranks 7--25 at the mode-space bipartitions (grid-independent).
Despite these properties, production timing showed TT-AMEn to be more than an order of magnitude slower than dense mode-parallel sweeps at device scale.

The root cause is structural and general. The streaming operator $\mathbf{v}_\lambda \cdot \nabla e_\lambda(\mathbf{r})$ is diagonal in mode space: for each $(\lambda, \theta, \varphi)$, it reduces to a scalar advection equation on the spatial grid, solvable by a single upwind sweep
in $O(N_{\mathrm{spatial}})$. Dense sweeps exploit this directly. TT-AMEn couples all modes through the TT ranks at every site, paying $O(r^2 n_k r_A^2)$ per site per sweep even when no inter-mode coupling is required for streaming. The local operator formation (59\% of sweep time) and dense local solve (37\%) dominate, with the ratio TT/dense growing as the grid is refined. The TT-AMEn experiment establishes that the BTE solution lies on a ${\sim}25$-dimensional manifold in joint mode--spatial space (TT bond rank at the mode-spatial bipartition saturates at 25 regardless of truncation threshold for the $N = 5$ run), a result we quantify in the next section.

\subsection{Part II: Solution manifold structure}\label{sec:partII}

In Part I, we showed the global incompressibility of the scattering matrix.  We now show an entirely different characteristic of the solution manifold.

\subsubsection{The non-equilibrium distribution is low rank}\label{sec:rank2}

We first establish the rank-2 solution structure in the 1D slab geometry and then confirm the analogous result in 3D.  The total phonon distribution $e_\lambda(x)$ is dominated by the local-equilibrium component $c_\lambda T(x)$ --- a rank-1 field. The physically interesting part is the non-equilibrium deviation $\delta e_\lambda(x) \equiv e_\lambda(x) - c_\lambda T(x)$. We compute $\delta e$ at $N = 5$ for converged full-$W$ solves in a 1D slab at thicknesses $L = 20, 40, 100, 200, 500$~nm.
The non-equilibrium fraction $\|\delta e\|_F / \|e\|_F$ varies from 0.1\% at $L = 20$~nm to 0.02\% at $L = 500$~nm.

The SVD of the matrix $[\delta e]_{n,x}$ (shape $747 \times 80$) yields the singular value structure shown in Table~\ref{tab:manifold}. The result is striking: $r(99\%) = 2$ at all slab thicknesses, from deeply ballistic to quasi-diffusive. The participation ratio $\mathrm{PR} \approx 0.022$ means only 2\% of the 747 modes carry 98\% of the non-equilibrium variance.

\begin{table}[ht]
\caption{Solution manifold rank for the 1D slab geometry: non-equilibrium deviation $\delta e = e - cT$ and beyond-RTA correction $\Delta e = E_{\mathrm{fW}} - E_{\mathrm{RTA}}$ ($N=5$, $r_{\mathrm{SVD}}=50$). All conductivities in \si{\watt\per\metre\per\kelvin}. Kn is conductivity-weighted mean Knudsen number.}\label{tab:manifold}
\begin{tabular}{@{}rrlllllll@{}}
\toprule
$L$ (nm) & Kn & 
  $r(99\%,\delta e)$ & $\mathrm{PR}(\delta e)$ & $r(99\%,\Delta e)$ \\
\midrule
 20 & ${\approx}\,7$   &  2 & 0.022 & 2 \\
 40 & ${\approx}\,3$   &  2 & 0.022 & 3 \\
100 & ${\approx}\,1$   &  2 & 0.022 & 4 \\
200 & ${\approx}\,0.5$ &  2 & 0.022 & 4 \\
500 & ${\approx}\,0.2$ &  2 & 0.020 & 4 \\
\botrule
\end{tabular}
\end{table}

The physical origin of rank-2 is transparent in the 1D geometry. The BTE solution for a mode pair $(\lambda, \bar\lambda)$ with $v_{\lambda,x} = v$ and $v_{\bar\lambda,x} = -v$ decomposes into symmetric and antisymmetric parts. The antisymmetric part $f^+ - f^-$ is the heat flux mode --- it is proportional to $v_{\lambda,x} \cdot (dT/dx)$ at leading order and constitutes singular vector~1. The symmetric correction from scattering constitutes singular vector~2. These two basis vectors span the entire non-equilibrium content regardless of how many phonon modes are present.

The column $r(99\%, \Delta e)$ in Table~\ref{tab:manifold} shows the rank of the \textit{beyond-RTA correction} $\Delta e = E_{\mathrm{fW}} - E_{\mathrm{RTA}}$. This rank stabilises at 4 for $L \geq 100$~nm (the quasi-diffusive regime). That is, the beyond-RTA correction is 4-dimensional in mode space.

Turning now to the 3D FinFET-like geometry, we measure the mode-space rank of the non-equilibrium distribution directly via the TT-AMEn experiment. The TT bond rank at the mode-spatial bipartition --- bond~2 (0-indexed) of the six-core TT chain $(\lambda, \theta, \varphi, x, y, z)$, which separates all mode and angular indices from all spatial indices --- equals by definition the numerical rank of the matrix in Eq.~(\ref{eq:unfolding}):

\begin{equation}
  \mathbf{e}_{(\lambda\theta\varphi),(xyz)}
    \in \mathbb{R}^{\Ma N_\Omega \times N_{\mathrm{cells}}}.
  \label{eq:unfolding}
\end{equation}

For the production $N = 5$ run ($\Ma = 747$, 128 angular directions), Supplementary Table~S4 reports this bond rank as \textbf{25} for the full distribution $e$, saturating regardless of TT truncation threshold (Supplementary Section~S6). The non-equilibrium deviation $\delta e = e - c_\lambda T(\mathbf{r})$ has mode-space rank $\leq 24$. Both values are larger than those for the 1D slab geometry due to additional degrees of freedom available in 3D but are still more than an order of magnitude below $\Ma = 747$ ($N = 5$) ---and more than two orders of magnitude below the production grid size $\Ma = 13{,}179$ ($N = 13$)--- confirming that the low-dimensional mode-space structure persists even in full 3D geometry with a localised volumetric heat source, corner effects, and multi-directional temperature gradients.

\subsection{Part III: Transport selectivity reconciles incompressibility with
  accuracy}\label{sec:partIII}

The fact that the non-equilibrium distribution has a low rank has a profound implication.  Despite the incompressibility of $\Win$ (requiring 89\% of rank globally) discussed in Part I, a low-rank BTE solver can still provide accurate solutions. As an example, for the 1D slab, our rank-50 BTE solver achieved 0.77\%  error in the effective conductivity at $N$ = 5 (Supplementary Table~S3).  The solution manifold has rank 4, and the leading 50 SVD modes of $\Win$ span this rank-4 subspace. The remaining almost 90\% of the SVD spectrum, while large in Frobenius norm, acts entirely within the local-equilibrium subspace and has no effect on transport observables.

We quantify this through the transport selectivity
\begin{equation}
  S = \frac{\|\Win - \WinR\|_F / \|\Win\|_F}{|\Delta k^{(r)}| / k_{\mathrm{ref}}},
  \label{eq:selectivity}
\end{equation}
where $\Delta k^{(r)} = k_{\mathrm{ref}} - k^{(r)}$ is the conductivity error at
rank $r$ relative to the near-full-rank reference.
$S \gg 1$ means the SVD truncation discards modes that are large in Frobenius norm
but transport-inert.
Table~\ref{tab:selectivity} shows results at all four BZ grid sizes.

\begin{table}[ht]
\caption{Transport selectivity $S$ across BZ grid sizes and ranks for the 1D slab geometry
  ($L = 100$~nm, $N_x = 80$). The transport observable is the effective thermal conductivity.  All thermal conductivities in  \si{\watt\per\metre\per\kelvin}.}\label{tab:selectivity}
\begin{tabular}{@{}rrrrrrrrrr@{}}
\toprule
$N$ & $\Ma$ & $\kRTA$ & $\kdense$ &
  $S(r{=}10)$ & $S(r{=}30)$ & $S(r{=}50)$ & $S(r{=}100)$ & $S(r{=}200)$ \\
\midrule
3 &    159 & 32.52 & 34.63 & 22.6 & 58.3 &  54.5 & 118.8 &  --- \\
5 &    747 & 73.38 & 76.65 & 35.7 & 76.0 &  60.8 &  36.8 & 48.6 \\
7 &  2,055 & 78.05 & 84.13 & 17.8 & 68.2 &  47.5 &  66.4 & 92.9 \\
9 &  4,371 & 74.71 & 80.78 & 18.5 & 96.0 &  85.1 & 108.9 & 384.8 \\
\botrule
\end{tabular}
\end{table}

\begin{figure}[t]
\centering
\includegraphics{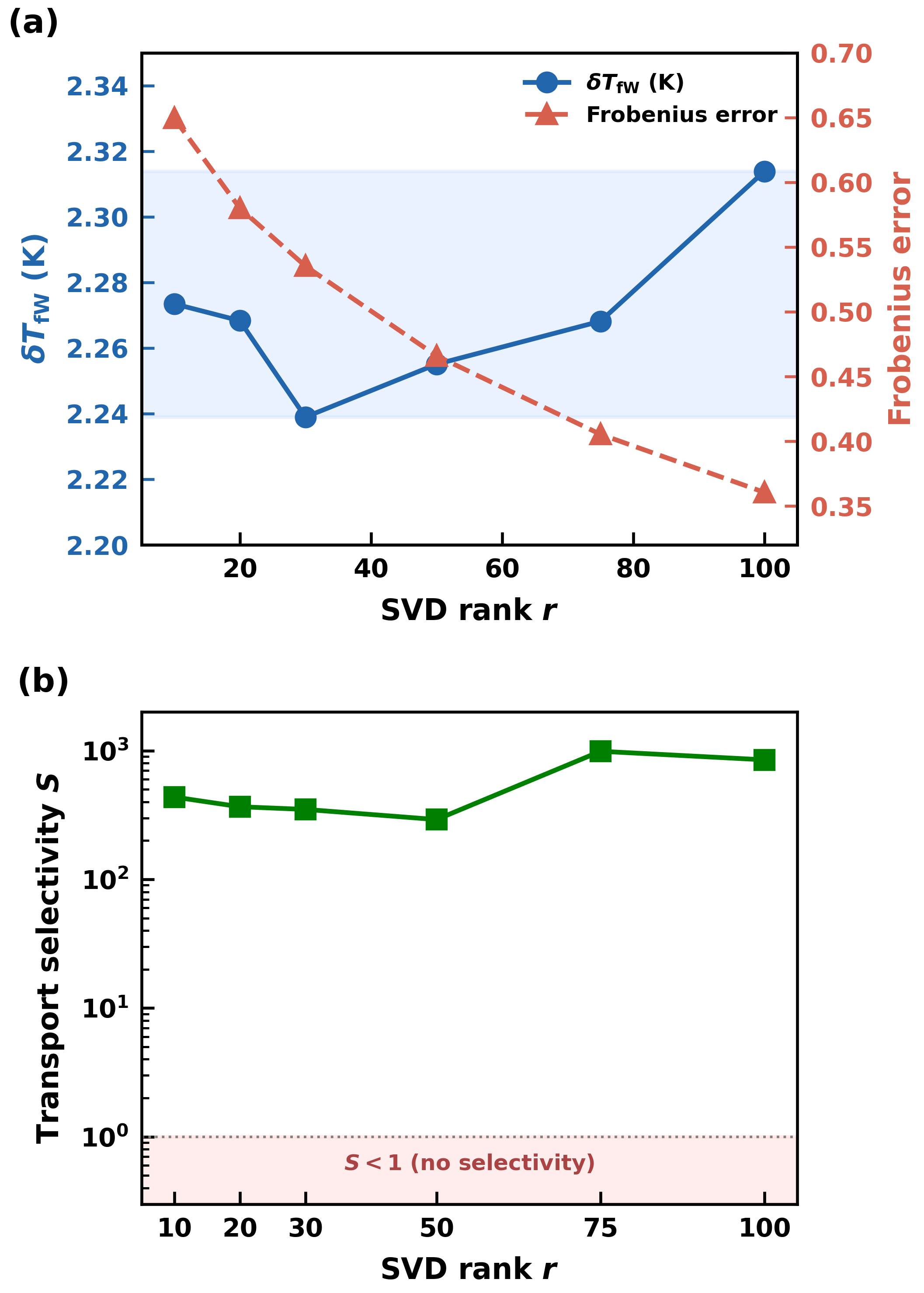}
\caption{Transport selectivity of the SVD truncation ($N=5$, FinFET).
  (a) $\dTfW$ (blue, left axis) and Frobenius error (red, right axis) vs.\ SVD
  rank $r$. Blue band: 3.3\% spread in $\delta T$; Frobenius error varies
  36--65\%.
  (b) Transport selectivity $S$ vs.\ $r$ for the 3D FinFET at $N=5$
  (Table~S6); values 55--120$\times$ throughout.}\label{fig:selectivity_finfet}
\end{figure}

Three observations follow from Table~\ref{tab:selectivity}. First, $S \gg 1$ at all $N$ and all tested ranks. Second, $S$ grows with $N$: at $r = 200$, $S = 93$ at $N=7$ vs.\ $S = 385$ at $N=9$.
Third, $S$ is non-monotone in $r$; the minimum transport error typically occurs in the range $r = 30$--100 depending on $N$. The rank $r = 50$ used throughout this study is within this range for all $N$.  

Figure~\ref{fig:selectivity_finfet} shows similar results for the 3D FinFET-like structure.  The correction
$\dTfW(N)$ is defined as the difference between the prediction under RTA and the prediction with the full scattering matrix: $\Trta(N) - \TfW(N)$.

\subsubsection{Why the streaming-diagonal structure is the enabling architecture}\label{sec:streaming}

The streaming operator $\mathbf{v}_\lambda \cdot \nabla$ is diagonal in mode space: for each mode $\lambda$, it acts on $e_\lambda(\mathbf{r})$ independently without coupling to other modes. Our hybrid architecture --- dense spatial sweeps for streaming + low-rank $\Win$ for scattering --- exploits the streaming diagonality where it exists and the solution
manifold compressibility where that exists.

This also explains why the TT-AMEn approach failed despite the solution lying on a 25-dimensional manifold: TT-AMEn applied low-rank structure to the streaming step (where mode-diagonality means rank~1 is exactly sufficient) rather than to the scattering step (where the rank-4 non-equilibrium structure provides genuine compression).

\subsection{Part IV: Solver validation and physics results}\label{sec:partIV}

\subsubsection{Equilibrium invariant}\label{sec:equil}

With uniform boundary conditions at $T_0 = \SI{300}{\kelvin}$ and no heat source,
the solver reproduces the exact solution $e_\lambda(\mathbf{r}) = 0$ to
$\max_i |T_i - T_0| < \SI{e-9}{\kelvin}$ at all BZ grid sizes.

\subsubsection{1D slab: analytic validation}\label{sec:1dslab}

Applying isothermal-diffuse wall conditions yields the per-mode effective conductivity
ratio
\begin{equation}
  \frac{k_{\mathrm{eff},n}}{k_n^x} = \frac{1}{1 + 2\,\mathrm{Kn}_n},\qquad
  k_n^x = \frac{c_n v_{n,x}^2\tau_n}{n_q V_{\mathrm{uc}}},\qquad
  \mathrm{Kn}_n = \frac{|v_{n,x}|\tau_n}{L},
  \label{eq:keff_mode}
\end{equation}
and the analytic IMA reference
\begin{equation}
  k_{\mathrm{eff}}^{\mathrm{Sond}} = \sum_n \frac{k_n^x}{1 + 2\,\mathrm{Kn}_n}.
  \label{eq:keff_sond}
\end{equation}
For cubic Si, $\sum_n k_n^x = \kbulk = \SI{148.0}{\watt\per\metre\per\kelvin}$
($< 0.003\%$ deviation by isotropy).

\begin{figure}[t]
\centering
\includegraphics{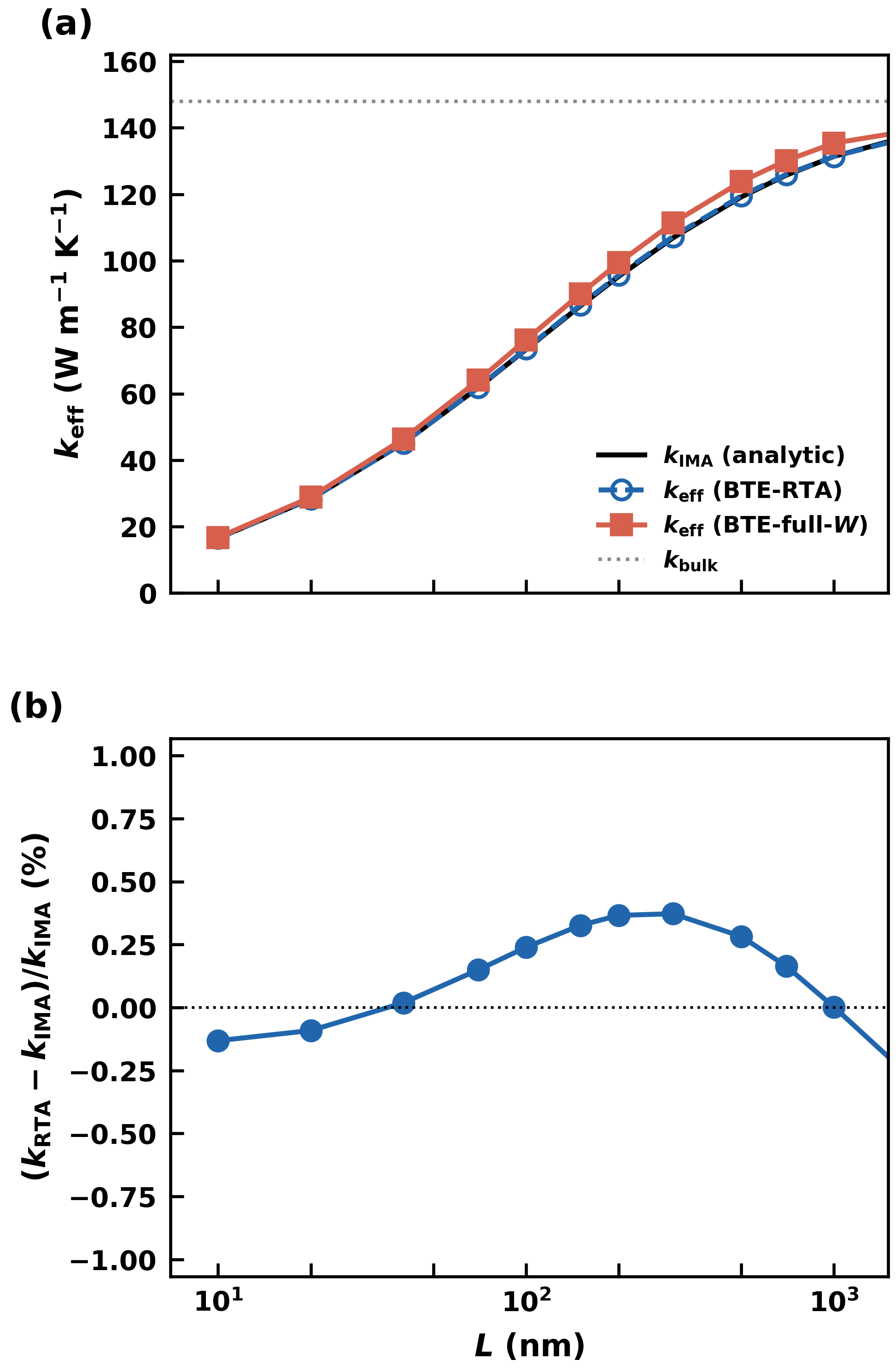}
\caption{Validation: 1D slab BTE vs.\ exact analytic IMA reference.
  (a) Effective thermal conductivity $\keff$ vs.\ slab thickness $L$, log-linear axes.
  Filled red squares: BTE full-$W$ solver (SVD $r=50$).
  Open blue circles: BTE-RTA solver ($N=5$, $N_x=100$).
  Solid black line: analytic IMA reference Eq.~(\ref{eq:keff_sond}).
  Dotted grey: $\kbulk = \SI{148.0}{\watt\per\metre\per\kelvin}$.
  (b) Relative deviation $(k_{\mathrm{RTA}} - k^{\mathrm{Sond}})/k^{\mathrm{Sond}}$
  vs.\ $L$. All values $\leq 0.37\%$ over $L = 10$--$1000$~nm.}\label{fig:1dslab}
\end{figure}

Table~\ref{tab:1dslab} and Fig.~\ref{fig:1dslab} compare $\keff$ from the solver and Eq.~(\ref{eq:keff_sond}) over $L = 10$--$1000$~nm. The maximum deviation between the RTA solver and the IMA is \textbf{0.37\%} (at $L = 200$--300~nm). Under full $W$, $\keff$ is 1.2--3.9\% higher than the RTA across all converged $L$, consistent with the later results from the FinFET-like geometry.

\begin{table}[ht]
\caption{1D slab validation: BTE solver vs.\ analytic IMA ($N=5$, $N_x=100$).
  Values in \si{\watt\per\metre\per\kelvin}.
  $\delta k_{\mathrm{fW}}\% = (k_{\mathrm{fW}} - \kRTA)/\kRTA$.}\label{tab:1dslab}
\begin{tabular}{@{}rrlllll@{}}
\toprule
$L$ (nm) & $\mathrm{Kn}_x$ & $k^{\mathrm{Sond}}$ & $\kRTA$ & err\%
  & $k_{\mathrm{fW}}$ & $\delta k_{\mathrm{fW}}$\% \\
\midrule
   10 & 6.73 & 16.526 & 16.504 & $-0.13$ & 16.704 & $+1.21$ \\
   40 & 1.68 & 45.139 & 45.148 & $+0.02$ & 46.418 & $+2.81$ \\
  100 & 0.67 & 73.304 & 73.479 & $+0.24$ & 76.171 & $+3.66$ \\
  200 & 0.34 & 95.299 & 95.647 & $+0.37$ & 99.382 & $+3.90$ \\
  500 & 0.14 & 119.199 & 119.535 & $+0.28$ & 123.803 & $+3.57$ \\
 1000 & 0.07 & 131.387 & 131.390 & $0.00$  & 135.392 & $+3.05$ \\
\botrule
\end{tabular}
\end{table}

\textbf{SVD rank sensitivity.}
$\dTfW$ varies by only 3.3\% relative across ranks $r = 10$--100, while the Frobenius
error varies from 65\% to 36\% (Fig.~\ref{fig:selectivity_finfet}).
Transport selectivity $S = 55$--$120\times$ across this range.

\textbf{DSA (Diffusion Synthetic Acceleration) fixed-point consistency.}
With and without DSA, $T_{\max}$ agrees to $\leq \SI{0.5}{\milli\kelvin}$ at all BZ
grid sizes.

\subsubsection{3D FinFET: BZ grid convergence}\label{sec:bzconv}

All computations use the FinFET-like geometry with $P = \SI{10}{\micro\watt}$
drain-side heat generation.
Table~\ref{tab:bzconv} presents the peak fin temperature under RTA ($\Trta$), under
full-$W$ ($\TfW$), the BZ-induced shift
$\Delta T_{\mathrm{BZ}}(N) = \Trta(N) - \Trta(5)$, and the correction
$\dTfW(N) = \Trta(N) - \TfW(N)$.

\begin{table}[ht]
\caption{BZ grid convergence of the full-$W$ correction.
  Ratio $= \dTfW / (\Trta - 300)$.
  DSA-confirmed: $\TfW(N{=}13) = \SI{314.717}{\kelvin}$,
  $\TfW(N{=}15) = \SI{314.826}{\kelvin}$
  ($\leq\SI{0.5}{\milli\kelvin}$ agreement).}\label{tab:bzconv}
\begin{tabular}{@{}rrllllr@{}}
\toprule
$N$ & $M$ & $\Trta$ (K) & $\TfW$ (K) & $\Delta T_{\mathrm{BZ}}$ (K)
  & $\dTfW$ (K) & Ratio \\
\midrule
 5 &    747 & 322.399 & 320.144 &  $0.000$ & 2.255 & 10.07\% \\
 7 &  2,055 & 318.202 & 315.954 & $-4.197$ & 2.248 & 12.35\% \\
 9 &  4,371 & 319.790 & 317.297 & $-2.609$ & 2.494 & 12.60\% \\
11 &  7,983 & 317.047 & 315.192 & $-5.352$ & 1.855 & 10.88\% \\
13 & 13,179 & 316.640 & 314.718 & $-5.759$ & 1.923 & 11.55\% \\
15 & 20,247 & 316.645 & 314.826 & $-5.754$ & 1.819 & 10.93\% \\
\botrule
\end{tabular}
\end{table}

\begin{figure}[t]
\centering
\includegraphics{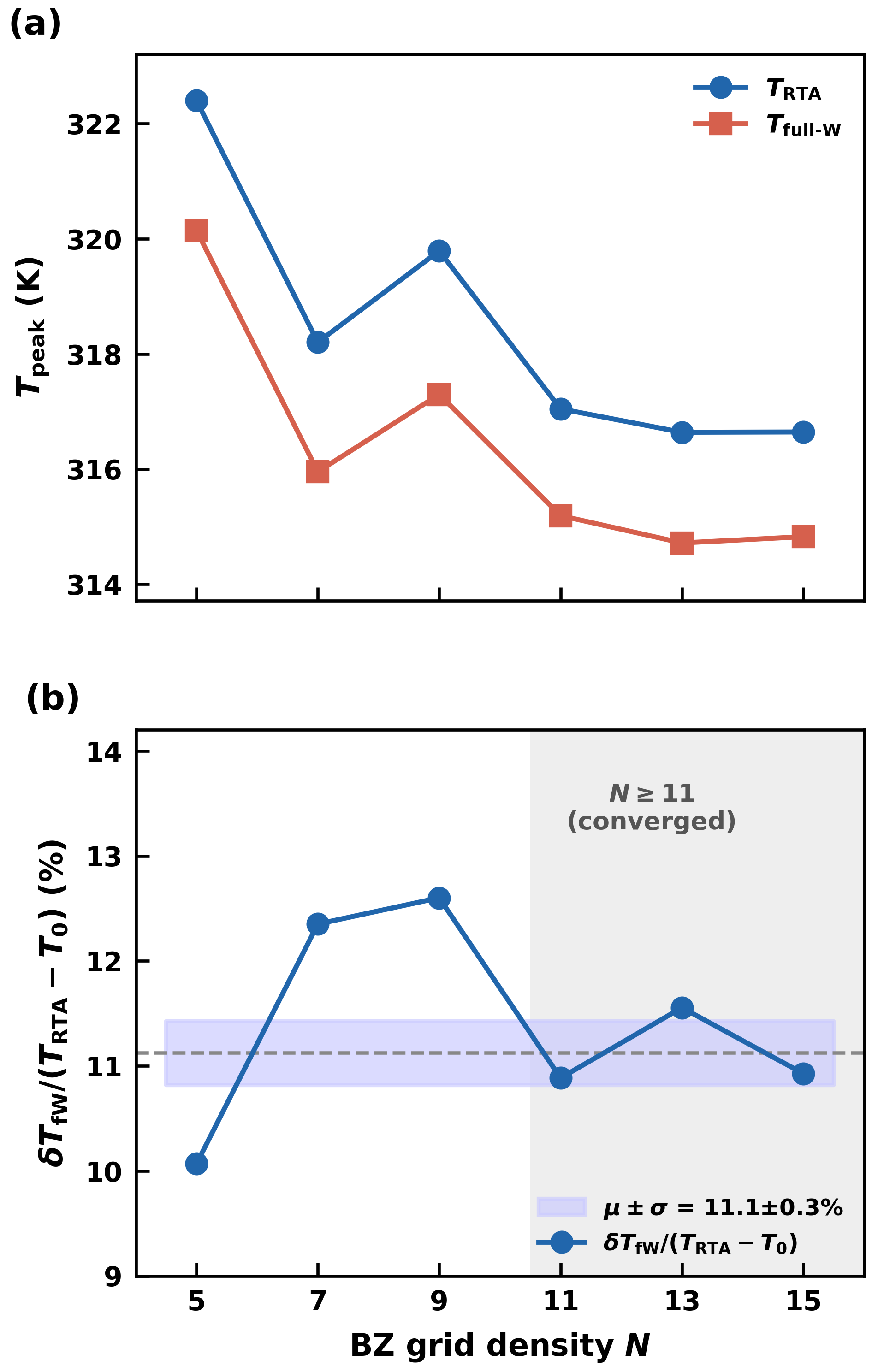}
\caption{Brillouin zone grid convergence.
  (a) $\Trta(N)$ (blue circles) and $\TfW(N)$ (red squares) vs.\ BZ grid density
  $N$.
  (b) Normalised correction $\dTfW/(\Trta - T_0)$ vs.\ $N$; grey band marks the
  asymptote $11.1 \pm 0.3\%$ ($N \geq 11$).
  The non-monotone behaviour at $N = 7, 9$ reflects the $\Gamma$-centred
  Monkhorst--Pack sampling artefact.}\label{fig:bzconv}
\end{figure}

The absolute temperatures $\Trta(N)$ exhibit non-monotone convergence at $N = 7, 9$ --- a known artefact of $\Gamma$-centred Monkhorst--Pack sampling: at odd $N$, the grid does not close under $\mathbf{q} \to -\mathbf{q}$, producing disjoint sampling of long-MFP acoustic branches \cite{Broido2007,Carrete2017}. The correction $\dTfW(N)$ is substantially less sensitive, because it is the difference of two solutions on the \textit{same} grid and the sampling error cancels to first order. Convergence of $\dTfW$ is achieved at $N \geq 11$ (Fig.~\ref{fig:bzconv}). Taking $N = 11, 13, 15$:
\begin{equation}
  \dTfW = 1.87 \pm 0.05\,\mathrm{K}
    = 11.2 \pm 0.3\%\text{ of the RTA temperature rise.}
  \label{eq:result}
\end{equation}

\subsubsection{3D FinFET: Ballistic invariance}\label{sec:ballistic}

\begin{figure}[t]
\centering
\includegraphics{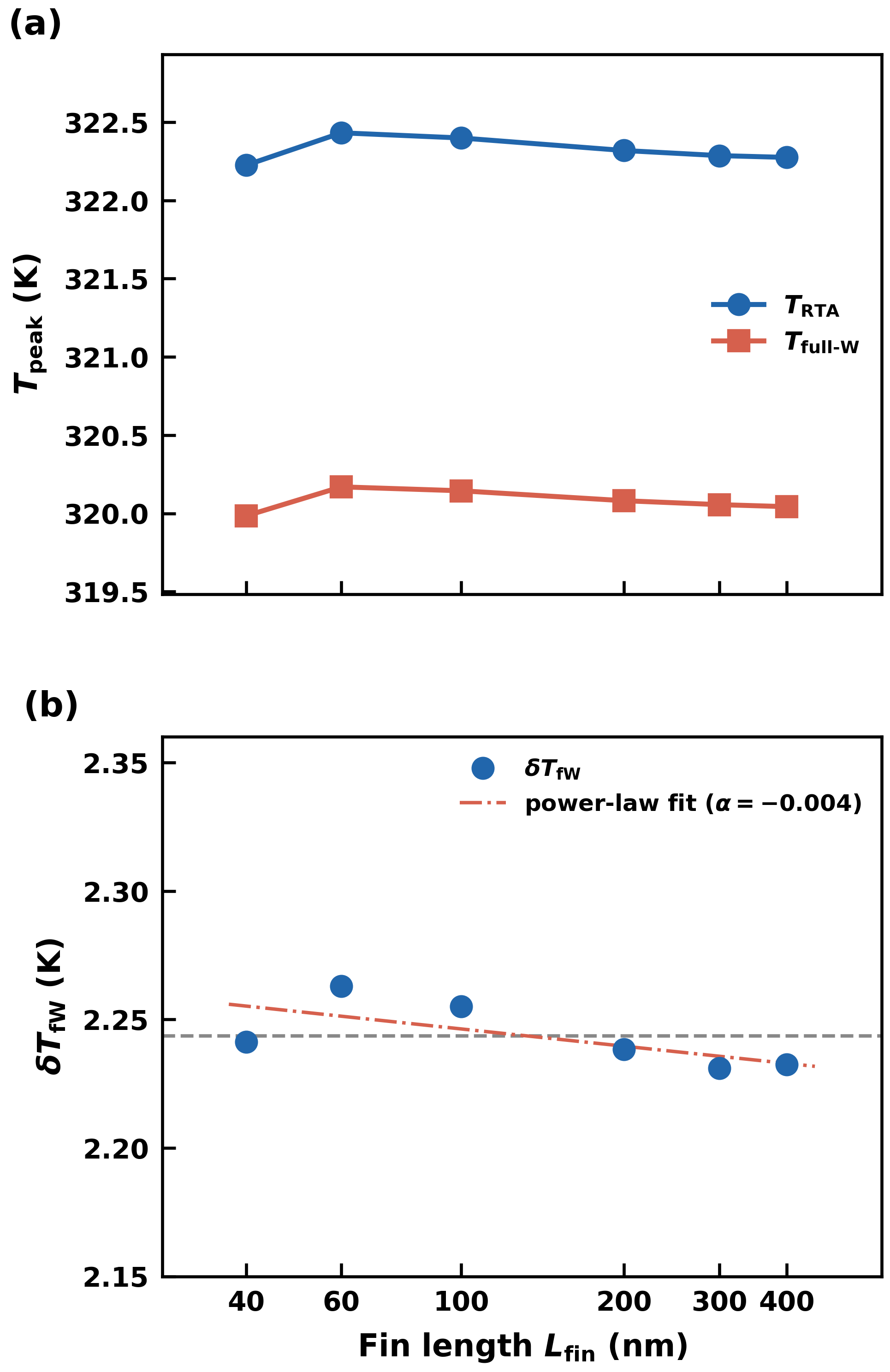}
\caption{Ballistic invariance of the full-$W$ correction.
  (a) Peak temperatures $\Trta$ and $\TfW$ vs.\ $L_{\mathrm{fin}}$. Both series are
  flat to within \SI{0.2}{\kelvin} across a decade of device lengths.
  (b) Absolute correction $|\dTfW|$ vs.\ $L_{\mathrm{fin}}$ with power-law fit
  ($\alpha = -0.004$). Grey band: $\pm 1\sigma$.}\label{fig:ballistic}
\end{figure}

Varying $L_{\mathrm{fin}} = 40$--400~nm at fixed $P = \SI{10}{\micro\watt}$ and $N = 5$ gives the results shown in Table~\ref{tab:ballistic} and Fig.~\ref{fig:ballistic}. A power-law fit gives $\alpha = -0.004 \approx 0$ ($R^2 = 0.48$, fin length explains $< 0.5\%$ of variance):
\begin{equation}
  |\dTfW|/\Delta T_{\mathrm{rise}} = 10.05 \pm 0.03\%,\qquad \alpha \approx 0.
  \label{eq:ballistic}
\end{equation}
This ballistic invariance is a consequence of the low-rank solution structure: the non-equilibrium deviation $\delta e$ may be approximated as occupying a rank-2 subspace determined by the phonon dispersion and scattering physics, not by the device geometry.

\begin{table}[ht]
\caption{Full-$W$ correction vs.\ fin length.}\label{tab:ballistic}
\begin{tabular}{@{}rlllr@{}}
\toprule
$L_{\mathrm{fin}}$ (nm) & $\Trta$ (K) & $\TfW$ (K) & $|\dTfW|$ (K) & Ratio \\
\midrule
 40 & 322.225 & 319.984 & 2.241 & 10.08\% \\
 60 & 322.432 & 320.169 & 2.263 & 10.09\% \\
100 & 322.399 & 320.144 & 2.255 & 10.07\% \\
200 & 322.319 & 320.081 & 2.238 & 10.03\% \\
300 & 322.287 & 320.056 & 2.231 & 10.01\% \\
400 & 322.275 & 320.043 & 2.232 & 10.02\% \\
\botrule
\end{tabular}
\end{table}

\subsubsection{Spatial structure}\label{sec:spatial}

\begin{figure}[t]
\centering
\includegraphics{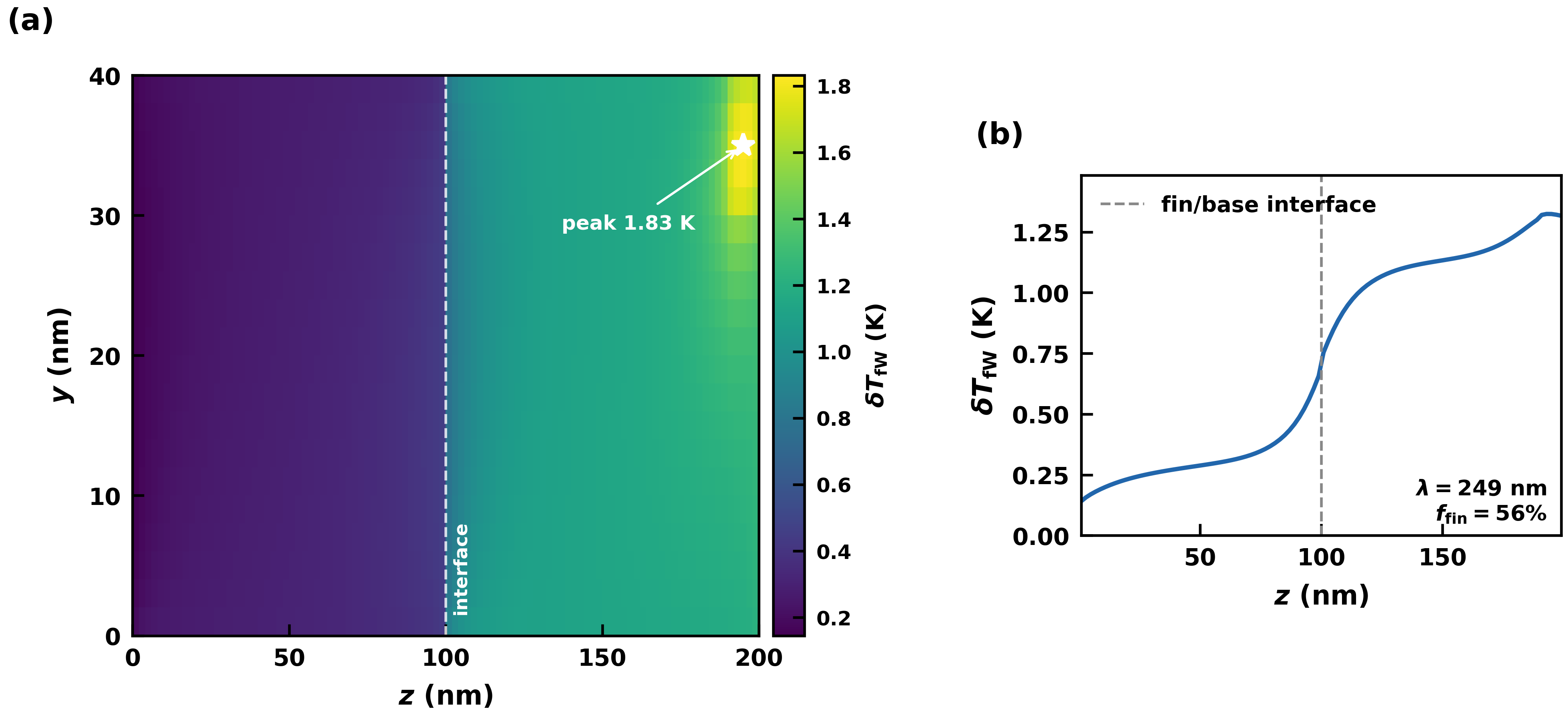}
\caption{Spatial structure of the full-$W$ correction field ($N=13$).
  (a) 2D heatmap $\delta T(y,z)$ at fin $x$-midplane. White star: peak
  \SI{1.92}{\kelvin} at drain-side apex.
  (b) Centreline $\delta T(z)$ vs.\ $z$; vertical dashed: fin--base interface;
  $\lambda = 249$~nm, $f_{\mathrm{fin}} = 56\%$ retention.}\label{fig:spatial}
\end{figure}

At $N = 13$, peak $\delta T_{\max} = \SI{1.923}{\kelvin}$ occurs at the drain-side
fin apex. Source concentration ratio is $1.54\times$, spatial decay length
$\lambda = 249$~nm $= 2.5\times$ the fin height, 57\% retention at the fin--base
interface (Fig.~\ref{fig:spatial}).

\section{Discussion}\label{sec:discussion}

\textbf{Physical interpretation of the correction.}
The full $W$ introduces inter-mode energy redistribution: Normal processes transfer energy between branches without resistance, while Umklapp processes couple modes with different mean free paths, allowing long-MFP modes to scatter more efficiently into short-MFP modes. The net effect --- for the SW phonon model --- is an $\sim$11\% reduction in effective ballistic thermal resistance, uniform across device geometry. This 11\% is specific to the SW potential; with ab initio force constants the magnitude may differ, but the qualitative physics and the structural conclusions are potential model-independent.

\textbf{The incompressibility result in broader context.}
Our result that $\Win$ requires 87--91\% of full rank and worsens with refinement has direct implications for algorithms that attempt global low-rank compression of the phonon scattering operator. The relaxon picture \cite{Cepellotti2016,Simoncelli2020} achieves elegant results for bulk homogeneous systems; but the gapless spectrum we find ($|\lambda_1|/|\lambda_{\Ma}| \approx 0.003$) confirms there is no natural truncation of the relaxon sum for the transport problem in finite devices. Green's function approaches \cite{Chiloyan2021} have so far been restricted to unbounded homogeneous media. The incompressibility result is a feature of three-phonon physics --- the $O(N_q^6)$ channel count --- not of the choice of potential or approximation level.

\textbf{Why the solution is nonetheless low-dimensional.}
Although $\Win$ is globally incompressible, the BTE solution is rank-2 in the
non-equilibrium subspace. The BTE acts as a low-pass filter --- fast-relaxing modes equilibrate locally and contribute only to the temperature, not to the heat flux. In 3D, the TT-AMEn mode-spatial bipartition measurement (Supplementary Section~S6) confirms the analogous result empirically: the mode-space rank of the non-equilibrium deviation $\delta e$ is $\leq 24$ for the FinFET-like geometry at $N = 5$, far below $\Ma = 747$. The additional degrees of freedom relative to rank-2 reflect the 3D spatial boundary layers at all six walls and the fin--base interface, and the multiple independent heat-flux directions in the three-dimensional domain (see Supplementary Section~S4 subsection 'Connection to TT-AMEn').

\textbf{Choice of the Stillinger--Weber potential.}
The SW potential \cite{Stillinger1985} was chosen for three specific reasons.
(i)~Closed-form analytic expressions for FC2 and FC3 on any supercell enable exact
verification of energy conservation $\sum_{\lambda'} W_{\lambda\lambda'} c_{\lambda'} = 0$ to machine precision. (ii)~The relative scattering rates $W_{\lambda\lambda'}/|W_{\lambda\lambda}|$ are
$\alpha$-independent, where $\alpha$ is the fitted global timescale introduced to match the known bulk thermal conductivity of silicon. (iii)~The SW potential is unambiguous and openly available. 

We also note a regime boundary on the rank-2 result: in materials where Normal
phonon--phonon scattering conserves crystal momentum strongly --- the
phonon-hydrodynamic regime, realised in graphene near room temperature and some III--V
compounds at low temperature --- a second slow hydrodynamic mode raises the
mode-space manifold rank above 2.

The global incompressibility result is nevertheless unconditionally universal; only the rank-2 result is specific to the non-hydrodynamic regime, which still encompasses most device-relevant semiconductors at room temperature. Extension to DFT-derived force constants from ShengBTE\cite{ShengBTE} or ALAMODE\cite{ALAMODE}, or machine-learned potentials\cite{Batatia2022}, is possible given our solver architecture.

\textbf{Why the structural results do not depend on the choice of potential.}
The three structural findings --- $\Win$ incompressibility, rank-2 solution manifold,
and transport selectivity --- are consequences of the BTE's mathematical structure,
not of the specific phonon model. The $O(\Ma^2)$ channel count and gapless relaxon spectrum are properties of three-phonon physics that hold for any material and any potential.

\textbf{Practical implications for full-$W$ device BTE at production scale.}
For $N = 20$ ($\Ma \approx 50{,}000$): $\Win$ will require $\sim$83\% of full rank
for 1\% Frobenius accuracy. But the solution manifold will still be rank-2 to rank-4, and the transport selectivity will exceed $S > 1000\times$. A rank $r \approx 50$--100 SVD of $\Win$ will therefore achieve $< 1\%$ transport error at production BZ grids.

\textbf{On tensor-train methods.}
The full 6D TT-AMEn experiment established an empirical rank of $\sim$25 for the BTE
solution manifold in joint mode--spatial space (Supplementary Section~S6).
TT-AMEn is not the right vehicle for exploiting this compression because the streaming
operator is mode-diagonal.
The correct architecture compresses only $\Win$, leaving streaming as dense
mode-parallel sweeps.


\section{Methods}\label{sec:methods}

\subsection{Governing equation}\label{sec:gov}

We solve the linearised steady-state BTE in deviation form, writing
$e_\lambda(\mathbf{r}) = \hbar\omega_\lambda[f_\lambda(\mathbf{r}) -
f_\lambda^{\mathrm{eq}}(T_0)]$ for a small deviation from reference equilibrium at
$T_0 = \SI{300}{\kelvin}$:
\begin{equation}
  \mathbf{v}_\lambda \cdot \nabla e_\lambda(\mathbf{r})
    = -\sum_{\lambda'} W_{\lambda\lambda'} e_{\lambda'}(\mathbf{r})
    + \dot{Q}_\lambda(\mathbf{r}),
  \label{eq:bte_dev}
\end{equation}
with energy conservation $\sum_\lambda W_{\lambda\lambda'} c_{\lambda'} = 0$
(enforced analytically) and
$T(\mathbf{r}) = T_0 + \sum_\lambda e_\lambda / C_{\mathrm{tot}}$.

\subsection{Phonon model}\label{sec:phonon}

SW force constants \cite{Stillinger1985} on a $3\times3\times3$ supercell;
three-phonon matrix elements from Fermi's golden rule with Gaussian broadening
$\sigma = \SI{0.8}{\tera\hertz}$.
A single global timescale is fitted to obtain $\kbulk = \SI{148.0}{\watt\per\metre\per\kelvin}$.
The raw stored $W$ is converted to physical units (\si{\per\second}) via $W_{nn}^{\mathrm{physical}} = -1/\tau_n$ (Supplementary Section~S2). Resulting relaxation times span 1.86--327~ps. Energy conservation is enforced by a rank-1 null-space projection at operator
application time.

\subsection{Device geometry and heat source}\label{sec:geometry}

FinFET-like structure: fin $20\times40\times L_{\mathrm{fin}}$~nm atop base
$60\times40\times100$~nm.
Isothermal substrate at \SI{300}{\kelvin}; all side walls diffuse-adiabatic.
Heat source: $y \in [L_y-10, L_y]$~nm, $z \in [L_{\mathrm{fin}}-10,
L_{\mathrm{fin}}]$~nm, full $x$;
$\dot{Q} = \SI{5e18}{\watt\per\cubic\metre}$ ($P = \SI{10}{\micro\watt}$).
Full geometry specification is given in Supplementary Section~S8.

\subsection{SVD compression of the scattering operator}\label{sec:svd}

The in-scattering operator $\Win = W + \mathrm{diag}(1/\tau_\lambda)$ is compressed
by truncated SVD at rank $r = 50$. Energy conservation is maintained after compression by a rank-1 correction at each matvec. The Frobenius error at $r = 50$ is 47\%; transport selectivity $S \approx 48$--$85\times$ (Table~\ref{tab:selectivity}) keeps transport error $< 1\%$.

\subsection{Spatial discretisation, angular quadrature, and
  iteration}\label{sec:disc}

Structured FVM is implemented on a $10\times20\times50$ fin mesh plus $30\times20\times50$ base mesh; $N_\Omega = 128$ angular directions. We adopt the upwind octant sweeps and Anderson mixing depth $m = 5$ using temperature-space weights. Diffusion Synthetic Acceleration (DSA) reduces iteration counts by 15--20\% with $\leq \SI{0.5}{\milli\kelvin}$ change in $T_{\max}$; derivation is in Supplementary Section~S7.
Convergence criterion is $\max_i|T_i^{(k+1)} - T_i^{(k)}|/\max_i|T_i^{(k)}| < 10^{-7}$.
The full-$W$ solver incurs an overhead of approximately $2$--$2.5\times$ over RTA:
the SVD scattering matvec costs $2r\Ma N_{\mathrm{cells}} \approx 5.97 \times 10^9$~FLOP per iteration vs.\ the streaming cost $N_\Omega \Ma N_{\mathrm{cells}} \approx 7.65 \times 10^9$~FLOP,
giving a per-iteration ratio of $\approx 1.77\times$; the additional 30--70\% more iterations to convergence (Supplementary Table~S8) brings the total to $\approx 2.4\times$. This overhead is independent of $N$ at fixed $r$.

\subsection{1D slab analytic reference}\label{sec:1d_ref}

The suppression function $(1 + 2\,\mathrm{Kn}_n)^{-1}$ in Eq.~(\ref{eq:keff_mode}) is exact for any single mode pair $(v,-v)$ under RTA with diffuse boundary conditions.

\subsection{Solution manifold rank measurement}\label{sec:rank_meas}

After a converged BTE solve, the SVD of the non-equilibrium deviation matrix
$[\delta e]_{\lambda,x} = [e]_{\lambda,x} - c_\lambda T(x)$ gives the mode-space
manifold rank at fractional variance threshold $p$:
$r^* = \min\{r : \sum_{i=1}^r \sigma_i^2/\sum_i \sigma_i^2 \geq p\}$.
Full data are in Supplementary Section~S4.


\backmatter

\bmhead{Supplementary information}

Supplementary information is available for this paper.


\section*{Declarations}

\begin{itemize}
\item \textbf{Funding:} Not applicable.
\item \textbf{Competing interests:} The authors declare no competing interests.
\item \textbf{Data availability:} All relevant computed results are available in the supplementary information.
\item \textbf{Code availability:} The solver source code will be deposited on GitHub under a permissive open-source license upon acceptance.
\item \textbf{Author contribution:} Y.S.J. initiated and conducted the whole research and wrote the manuscript.
\end{itemize}


\bibliography{references}

\end{document}